\documentclass[prl,nofootinbib,superscriptaddress,twocolumn]{revtex4}
\usepackage[a4paper,left=1.5cm,right=1.5cm,top=3cm,bottom=3cm]{geometry}

\usepackage{amssymb,amsmath,amsfonts}
\usepackage[dvipsnames]{xcolor}
\usepackage{graphicx}
\usepackage{longtable}
\usepackage{verbatim}
\usepackage{color}
\usepackage{mdframed}
\usepackage{soul}
\usepackage{amsfonts,amssymb,mathrsfs,amsmath,esint}
\usepackage{slashed, cancel}
\usepackage{framed}
\usepackage{mdframed}
\usepackage{simplewick} % For Wick contractions
\allowdisplaybreaks

\usepackage{latexsym}
\usepackage{graphicx}
\graphicspath{{./Figures/}}
\usepackage[dvipsnames]{xcolor}
\usepackage{booktabs}
\usepackage{datetime}
\newdateformat{mydate}{\THEDAY{ }\monthname[\THEMONTH]{ }\THEYEAR}

\usepackage{tikz}
\usepackage{color}
\usepackage{framed}
\usepackage{hyperref}
\hypersetup{colorlinks, citecolor=bluscuro, linkcolor=black, urlcolor=bluscuro}
\definecolor{rossos}{cmyk}{0,1,1,0.55}
\definecolor{bluscuro}{rgb}{0.15, 0.2, .85}
\definecolor{bluchiaro}{cmyk}{1,.3,0.,0.1}

\graphicspath{{./Figures/}}

\newcommand{\be}{\begin{equation}}
\newcommand{\ee}{\end{equation}}
\newcommand{\bea}{\begin{eqnarray}}
\newcommand{\eea}{\end{eqnarray}}
\newcommand{\beq}{\begin{equation}}
\newcommand{\eeq}{\end{equation}}
\def\beqa{\begin{eqnarray}}

\def\vk{{\vec{k}}}
\def\vp{{\vec{p}}}

\def\d{{\rm d}}
\def\lisa{\text{\tiny LISA}}

\def\PBH{\text{\tiny PBH}}
\def\M{{\tiny M}}

\newcommand{\FF}[1]{\widetilde{#1}}

\newcommand{\Ic}{\mathcal{I}_c}
\newcommand{\Is}{\mathcal{I}_s}

\newcommand{\Pz}{\mathcal P_\zeta}

\newcommand{\OGW}{\Omega_\text{GW}}

\newcommand{\dd}{{\rm d}}
\def\eeqa{\end{eqnarray}}

\def\lsim{\mathrel{\rlap{\lower4pt\hbox{\hskip0.5pt$\sim$}}
    \raise1pt\hbox{$<$}}}         %less than or approx. symbol
\def\gsim{\mathrel{\rlap{\lower4pt\hbox{\hskip0.5pt$\sim$}}
    \raise1pt\hbox{$>$}}}         %greater than or approx. symbol

\newcommand{\half}{{\textstyle \frac{1}{2}}}

\newcommand{\quarter}{{\textstyle \frac{1}{4}}}

\usepackage[normalem]{ulem}

\newcommand{\arXiv}[2]{\href{http://arxiv.org/pdf/#1}{{\tt [#2/#1]}}}
\newcommand{\arXivold}[1]{\href{http://arxiv.org/pdf/#1}{{\tt [#1]}}}

\begin{document}

\title{The  Primordial Black Hole Dark Matter   - LISA Serendipity}
\author{N. Bartolo}
\address{Dipartimento di Fisica e Astronomia ``G. Galilei",
Universit\`a degli Studi di Padova, via Marzolo 8, I-35131 Padova, Italy}

\address{INFN, Sezione di Padova,
via Marzolo 8, I-35131 Padova, Italy}

\address{INAF - Osservatorio Astronomico di Padova, Vicolo dellOsservatorio 5, I-35122 Padova, Italy}

\author{V. De Luca}
\address{D\'epartement de Physique Th\'eorique and Centre for Astroparticle Physics (CAP), Universit\'e de Gen\`eve, 24 quai E. Ansermet, CH-1211 Geneva, Switzerland}

\author{G. Franciolini}
\address{D\'epartement de Physique Th\'eorique and Centre for Astroparticle Physics (CAP), Universit\'e de Gen\`eve, 24 quai E. Ansermet, CH-1211 Geneva, Switzerland}

\author{A. Lewis}
\address{Department of Physics and Astronomy, University of Sussex, Brighton BN1 9QH, U.K}

\author{M. Peloso}
\address{Dipartimento di Fisica e Astronomia ``G. Galilei",
Universit\`a degli Studi di Padova, via Marzolo 8, I-35131 Padova, Italy}

\address{INFN, Sezione di Padova,
via Marzolo 8, I-35131 Padova, Italy}

%\address{INAF - Osservatorio Astronomico di Padova, Vicolo dellOsservatorio 5, I-35122 Padova, Italy}

\author{A.~Riotto}
\address{D\'epartement de Physique Th\'eorique and Centre for Astroparticle Physics (CAP), Universit\'e de Gen\`eve, 24 quai E. Ansermet, CH-1211 Geneva, Switzerland}

\date{\today}

\begin{abstract}
\noindent
There has recently been renewed interest in the possibility that the dark matter in the universe consists of primordial black holes (PBHs). Current observational constraints leave only a few PBH mass ranges for this possibility.  One of them is around $10^{-12} M_\odot$. If PBHs with this mass are formed due to an enhanced scalar-perturbation amplitude, their formation  is  inevitably accompanied by the generation of gravitational waves (GWs) with frequency peaked in the mHz range, precisely around the maximum sensitivity of the LISA mission.
We show that, if these primordial black holes are the dark matter, LISA will be able to detect  the associated  GW power spectrum.
Although the GW source signal is intrinsically non-Gaussian, the  signal measured by LISA  is a sum of the signal from a large number of independent sources suppressing the non-Gaussianity at detection to an unobservable level. We also discuss the effect of the GW propagation in the perturbed universe.  PBH dark matter generically leads to a detectable, purely isotropic, Gaussian and unpolarised GW signal, a prediction that is testable with LISA.
\end{abstract}

\maketitle

\paragraph{Introduction.}

The existence and the nature of dark matter  remains one of the main puzzles in physics \cite{bertone}. The recent detection of GWs generated by the  merging of two $\sim 30\, M_\odot$ black holes \cite{ligo} has renewed the interest in the possibility that all (or a significant part of) the dark matter of the universe is in the form of PBHs (see Refs. \cite{bird,juan,revPBH,revPBH1} for  recent literature).

A standard way to generate PBHs in the early universe is to enhance the power spectrum of the  comoving curvature perturbation $\zeta$ during inflation, but only on scales much smaller than those constrained to be small by CMB observations \cite{s1,s2,s3} (see Ref. \cite{Espinosa:2017sgp} in the case in which  Standard Model Higgs perturbations are used).
After reheating the perturbations are transferred to the radiation, forming PBHs upon horizon re-entry if the perturbations are large enough.
A region typically collapses to a PBH at horizon entry if  the comoving density contrast during radiation domination $\Delta(\vec x)=4\nabla^2\zeta(\vec x)/(9a^2H^2)$ is larger than a critical value $\Delta_{\rm c}$
%\textcolor{red}{, which  is a function of  the shape of the power spectrum   \cite{musco}} 
(here $a$ is the scale factor and $H$ the Hubble parameter).

We define the comoving curvature perturbation power spectrum   as
\begin{equation}
\big<\zeta(\vk_1)\zeta(\vk_2)\big>' =\frac{2\pi^2}{k_1^3} \Pz(k_1),
\label{eq: def P zeta}
\end{equation}
where  we have adopted  the standard prime notation indicating that  we do not explicitly
write down the $(2\pi)^3$ times the Dirac delta of momentum conservation.
It is convenient to define the variance of $\Delta(\vec x)$ as
\be
\sigma^2_\Delta(M) =\int_0^\infty {\rm d}\ln k \,W^2(k,R_H){\cal P}_{\Delta}(k),
\ee
where we have  made use of a (gaussian) window function  $W(k,R_H)$ to  smooth out $\Delta(\vec x)$ on the comoving horizon length  $R_H\sim 1/aH$ and  ${\cal P}_{\Delta}(k)=(4k^2/9a^2H^2)^2
\Pz(k)$.
Assuming Gaussian primordial perturbations, the mass fraction $\beta_\M$ of the universe which ends up in PBHs at the time of formation  is approximately (for the non-Gaussian extension see Ref. \cite{ng})
\be
\label{beta}
\beta_\M=\int_{\Delta_{\rm c}}^\infty \frac{{\rm d}\Delta}{\sqrt{2\pi}\,\sigma_\Delta}e^{-\Delta^2/2\sigma_\Delta^2}\simeq\frac{\sigma_\Delta}{\Delta_{\rm c}\sqrt{2\pi}} e^{-\Delta_{\rm c}^2/2\sigma^2_\Delta}.
\ee
This  corresponds to a present fraction of dark matter $ f_\PBH(M)  \equiv  {\rm d} \, (\rho_\PBH/\rho_{\text{\tiny DM}})/{\rm d}  \ln M $ in form of  PBHs of mass $M$~\cite{revPBH}
\be
\label{f}
f_\PBH \left( M \right) \simeq\left(\frac{\beta_\M}{6\cdot 10^{-9}}\right)\left(\frac{\gamma}{0.2}\right)^{\half}\left(\frac{106.75}{g_*}\right)^{\quarter}\left(\frac{M_\odot}{M}\right)^{\half},
\ee
for a dark matter density parameter today $\Omega_{\rm DM}h^2\approx 0.12$.
Here $\gamma<1$ accounts for the efficiency of the collapse and $g_*$ is the number of effective relativistic degrees of freedom at horizon entry. We will take $\gamma \simeq 0.2$ \cite{Carr:1975qj}.

The key point is that, if there are large gradients in the curvature perturbations (for example
generated during the last stages of inflation), they inevitably act as a (second-order) source  \cite{Acquaviva:2002ud, Mollerach:2003nq, Ananda:2006af, Baumann:2007zm} of primordial GWs~\cite{saito,gbp}.
We can relate the mass $M$ of a PBH to the  peak frequency  of the GWs (not far from  the peak frequency of the corresponding curvature perturbations which collapse to form a PBH at horizon entry, $k= 2\pi f=aH$)~\cite{gbp}
\be
M\simeq 33\,\gamma \left(\frac{10^{-9}\,{\rm Hz}}{f}\right)^2 M_\odot.
\ee
\begin{figure}[h!]
\includegraphics[width=\columnwidth]{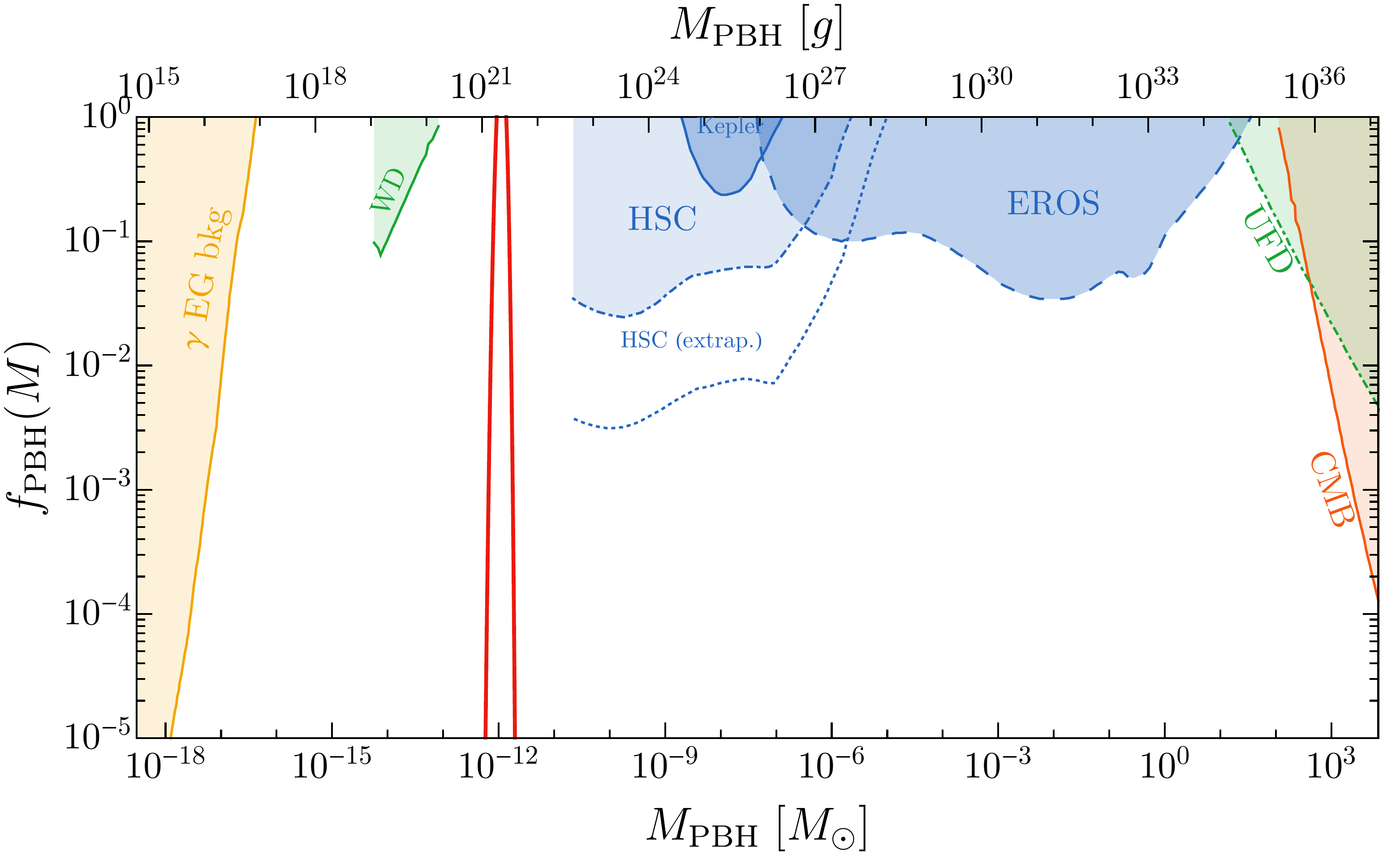}
\caption{Current  experimental constraints on monochromatic spectra of PBH at various masses (from Ref. \cite{japanese} and references therein).
 The PBH abundance shown as the red line (corresponding to all of the dark matter) has been obtained for $A_s=0.033$ and $k_\star=2\pi f_\lisa$ in Eq. (\ref{shape}).   }
\label{fig: Omega PBH}
\end{figure}
This shows that the mass corresponding to the frequency where  the Laser Interferometer Space Antenna (LISA) project  \cite{Audley:2017drz} has the maximum
sensitivity, $f_\lisa\simeq 3.4$ mHz,    is  $M\simeq  10^{-12} M_\odot$.

The serendipity is that around this mass
current observational constraints on the PBH abundances  are basically absent  \cite{k}, thus allowing $f_\PBH \left( M \right) \simeq 1$, see Fig. \ref{fig: Omega PBH}. Indeed,  the Subaru HSC  microlensing  measurements  \cite{hsc}    must be  cut around $10^{-11} M_\odot$, since
below this  mass the geometric optics approximation is no longer valid \cite{hsc1,k}:  the angular Einstein radius becomes much smaller than the angular size of the star, and the
magnification is then too small to be detected~\cite{hsc1,k}.
Neutron star limits \cite{neut} are also not included as they depend  on rather controversial  assumptions about the dark matter density in globular clusters \cite{k}.
The curious reader can find a more expanded discussion in Appendix A of Ref. \cite{long}.

It is an exciting coincidence that the optimal frequency range for the
LISA observatory corresponds to the mass range where PBHs can account for all the dark matter.
In this letter we show that, if dark matter is composed of PBHs of masses around $10^{-12} M_\odot$, then
LISA will measure the power spectrum of GWs inevitably associated with the production of the PBHs.
Furthermore, and despite the fact that the generated GWs are intrinsically non-Gaussian (their small-scale source is second-order in the curvature perturbation), we show that the signal measured by LISA would be highly Gaussian. This is because, as with other cosmological GW signals, a very large number of Hubble patches are observed over the resolution area of LISA, giving strong central limit theorem Gaussianisation~\cite{Caprini:2018mtu}. We also comment on GW propagation in the perturbed universe and primordial non-Gaussianity, neither of which affect the conclusion.

This short note contains only the main results; the reader can find the technical details  in Ref. \cite{long}.
\vskip 0.3cm
\noindent

\paragraph{PBHs as dark matter.}
\noindent
From Eq. \eqref{f} we see that PBHs of mass $\sim 10^{-12} M_\odot$ will form all of the dark matter if their corresponding mass fraction
is $\beta_\M\sim 6\cdot 10^{-15}$. As a benchmark example, we take  the comoving curvature perturbation power spectrum (augmented by the standard flat spectrum on large CMB scales)
to be the limiting case of a Dirac delta  function
\begin{eqnarray}
\label{shape}
{\cal P}_\zeta(k)&=&A_s\,k_\star\delta(k-k_\star).
\end{eqnarray}
Assuming this spectrum has the huge advantage that we can perform all the calculations analytically. Fig. \ref{fig: Omega PBH} shows the corresponding abundance of PBHs for a representative choice of parameters. We take $k_\star R_H\simeq 1$ and $\Delta_{\rm c}\simeq 0.45$. The   precise value of the threshold depends on the shape of the
power spectrum \cite{musco}, but this does not much alter the value of the spectrum amplitude $A_s$, which is the most relevant quantity for the amplitude of GWs produced.
The value of $A_s$ does depend on our assumption of Gaussian perturbations, which may well not be accurately valid since $A_s \sim 0.03$ is quite large. However, even if positive skewness of the $\Delta$ distribution meant that $f_\PBH\sim 1$ could be obtained with a lower $A_s$,
so that $\Delta_{\rm c}$ was then several more standard-deviation units away from zero, the required variance (proportional to $A_s$) would only change by an order unity factor (compared to the $\sim {\cal O}(100)$ reduction that would be required for the GW signal to become undetectable), so our conclusion should remain robust.

\vskip 0.3cm
\noindent

\begin{figure}[t!]
\includegraphics[width=\columnwidth]{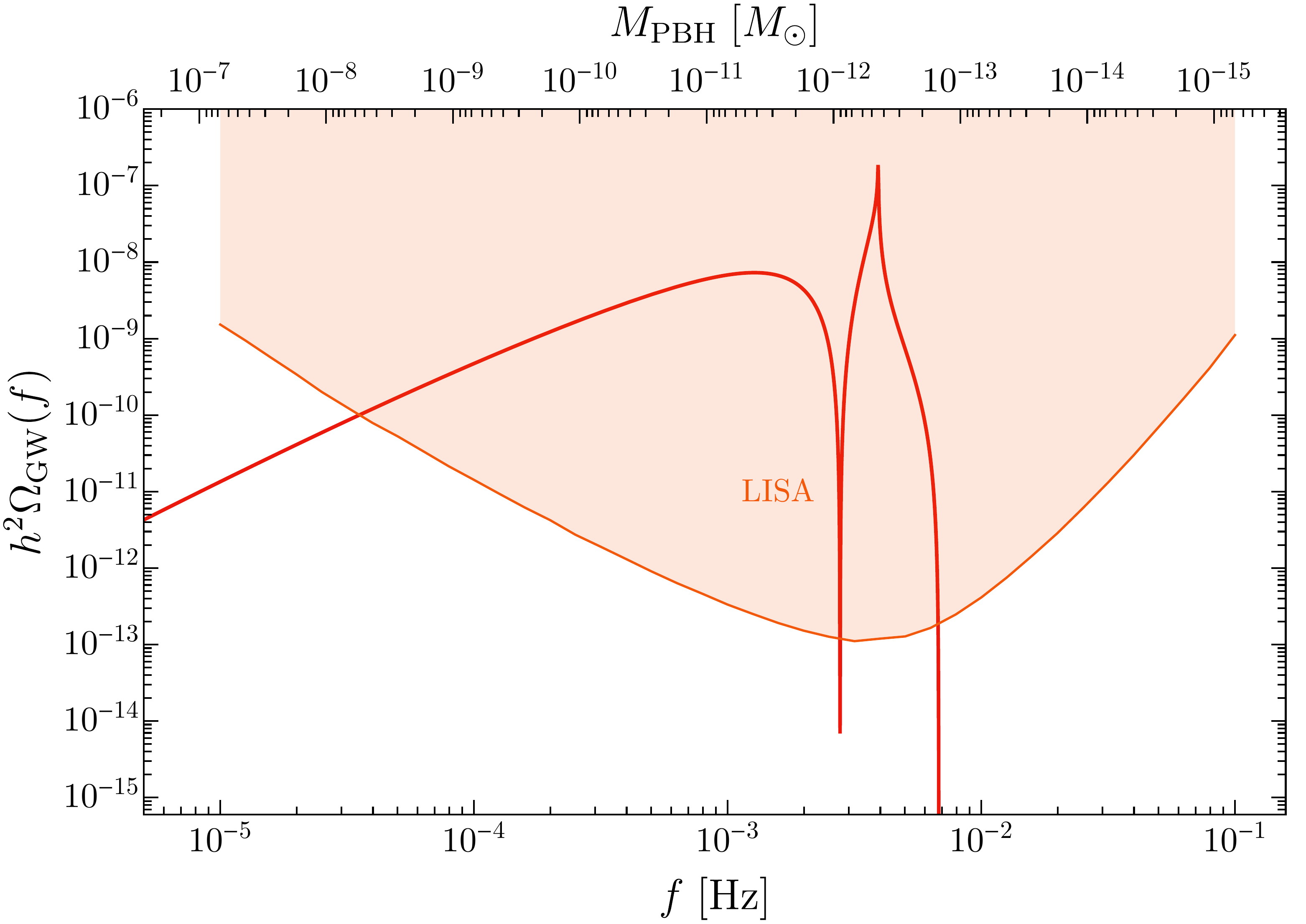}
\caption{The power spectrum of GWs  generated by PBHs compared with the power-law integrated sensitivity for LISA  estimated on the basis of the proposal \cite{Audley:2017drz}: the proposed design (4y, 2.5 Gm of length, 6 links) is  anticipated to have a sensitivity in between those called  C1 and C2 in  Ref. \cite{Caprini:2015zlo}. 
The spike is due to the trigonometric functions coming from  the radiation transfer functions in  $\mathcal{I}^2$, 
giving a resonant effect at $f \sim 2 f_{\rm LISA} / \sqrt{3}$, as explained in Ref. \cite{Ananda:2006af}.
The spike and slow fall-off in power to low frequencies are an artefact of assuming a monochromatic power spectrum; physical spectra would typically give a smooth spectrum with white-noise ($\propto f^3$) at low frequencies~\cite{errgw}, but a similar overall amplitude.
}
\label{fig: Omega GW}
\end{figure}

\paragraph{The power spectrum of GWs.}
We define the Newtonian-gauge scalar metric perturbation $\Psi$ and the transverse-traceless tensor metric perturbation $h_{ij}$ so that the linearized line element in tightly-coupled radiation domination is
\begin{equation}
 \d s^2\! =\! a^2\left\{-(1+2\Psi){\rm d}\eta^2 + \left[(1-2\Psi)\delta_{ij}+ \frac{h_{ij}}{2}\right]{\rm d}x^i {\rm d}x^j\right\}.
\end{equation}
We neglect the rare areas of strongly non-linear GW production associated directly with PBH formation and evolution, and focus on the signal sourced everywhere by second-order combinations of the linear scalar perturbations.
The equation of motion for the GWs is then obtained by expanding Einstein's equations up to second-order in the linear perturbations
\begin{equation}
h_{ij}''+2\mathcal H h_{ij}'-\nabla^2 h_{ij}=-4 \mathcal T_{ij}{}^{\ell m}\mathcal S_{\ell m},
\label{eq: eom GW1}
\end{equation}
where $'$ is the derivative with respect to the  conformal time $\eta$, $\mathcal H=a'/a$ is the conformal Hubble parameter
 and
 $\mathcal T_{ij}{}^{\ell m}$ projects the source term $\mathcal S_{\ell m}$ into its transverse and traceless part.
In the radiation phase the source is given by~\cite{Acquaviva:2002ud}
\begin{equation}
\label{psi}
\mathcal S_{ij}=2\partial_i\partial_j\left(\Psi^2\right)-2\partial_i\Psi\partial_j\Psi-\partial_i\left(\frac{\Psi'}{\mathcal H}+\Psi\right)\partial_j\left(\frac{\Psi'}{\mathcal H}+\Psi\right).
\end{equation}
Since this is  second-order in the perturbations, the sourced GWs are intrinsically non-Gaussian. The source is also local, depending only on spatial derivatives of the perturbations, so the resulting bispectrum will peak in momentum-space configurations where the wavevectors have similar amplitude (no squeezed component). 
 We define the projector in Fourier space using the chiral basis
\begin{equation}
\FF{\mathcal T}_{ij}{}^{\ell m}(\vk)=e_{ij}^{\rm L}(\vk)\otimes e^{{\rm L}\ell m}(\vk)
  + e_{ij}^{\rm R}(\vk)\otimes e^{{\rm R}\ell m}(\vk),
\end{equation}
where $e_{ij}^{\rm L,R}$ are the polarisation tensors.
In Eq. (\ref{psi}) the scalar perturbation $\Psi(\eta,\vk)$ can be written in terms of the  initial gauge-invariant comoving curvature perturbation as~\cite{lrreview}
\begin{equation}
\Psi(\eta,\vk) \equiv \frac 23 T(k\eta) \zeta(\vk) ,
\label{eq: Psi to zeta}
\end{equation}
where the transfer function during radiation domination with constant degrees of freedom is $T(x)= (9/x^2)\left[\sin (x/\sqrt 3)/(x/\sqrt 3) -\cos(x/\sqrt 3) \right]$. A straightforward calculation approximating the primordial perturbations as Gaussian leads to the current abundance of GWs \cite{errgw}
\begin{multline}
\frac{\OGW(f)}{\Omega_{r,0}} = \frac{c_g}{72}
  \int_{-\frac{1}{\sqrt{3}}}^{\frac{1}{\sqrt{3}}}\dd d \int_{\frac{1}{\sqrt{3}}}^{\infty}\hspace{-5pt}\dd s
  \left[ \frac{(d^2-1/3)(s^2-1/3)}{s^2-d^2} \right]^2\phantom{xxxx}\\
\cdot  \Pz\left(\frac{k\sqrt{3}}{2}(s+d)\right) \Pz\left(\frac{k\sqrt{3}}{2}(s-d)\right){\cal I}^2(d,s),
\label{eq: Omega GW with PS0}
\end{multline}
where $k=2\pi f$, $\Omega_{r,0}$ parameterises the current density of radiation if the neutrinos were massless, $c_g\simeq 0.4$ accounts for the change of the effective degrees of freedom of the thermal radiation during the evolution (assuming Standard Model physics), ${\cal I}^2\equiv\Ic^2+\Is^2$, and
\begin{eqnarray}
\Ic(x,y) &=&4  \int_0^\infty \dd\tau \, \tau (-\sin \tau)  \Big[ 2T(x\tau)T(y\tau) \nonumber\\
&+& \Big(T(x\tau)
+ x\tau\, T'(x\tau) \Big)\Big(T(y\tau) + y\tau\, T'(y\tau) \Big) \Big], \nonumber\\
&&
\label{eq: Ic, Is}
\end{eqnarray}
$\Is(x,y)$  being the same function, but with $\sin \tau$ replaced by $(-\cos\tau)$,
see Ref.~\cite{Kohri:2018awv}.
For the monochromatic power spectrum (Eq.~\eqref{shape}) we obtain (see also Refs. \cite{Ananda:2006af,saito,Kohri:2018awv})
\begin{align}
\frac{\OGW(f)}{\Omega_{r,0}} &=\frac{A_s^2 c_g f^2}{15552f_\star^2}\left(\frac{4f_\star^2}{f^2}-1\right)^2\hspace{-0.2cm}\theta\left(2-\frac{f}{f_\star}\right){\cal I}^2\left(\frac{f_\star}{f},\frac{f_\star}{f}\right),
\label{eq: Omega GW with PS}
\end{align}
where $f_\star=k_\star/2\pi$ and $\theta(x)$ is the step function.
The current abundance of GWs
is given in Fig. \ref{fig: Omega GW}  with $k_\star\sim k_\lisa=2\pi f_\lisa$ and $A_s\sim 0.033$.
Since the result is only a function of $f/f_\star$, for other possible $f_\star$ (with typical black hole masses as indicated on the top axis) the predicted spectrum simply shifts sideways in $f$.
This shows that, if PBHs of masses in the range $10^{-15} M_\odot \alt M\alt 10^{-11} M_\odot$ form the dark matter (or even a fraction of it), LISA will measure the GWs popping out during the PBH formation time.

\begin{figure}[h!]
\includegraphics[width=\columnwidth]{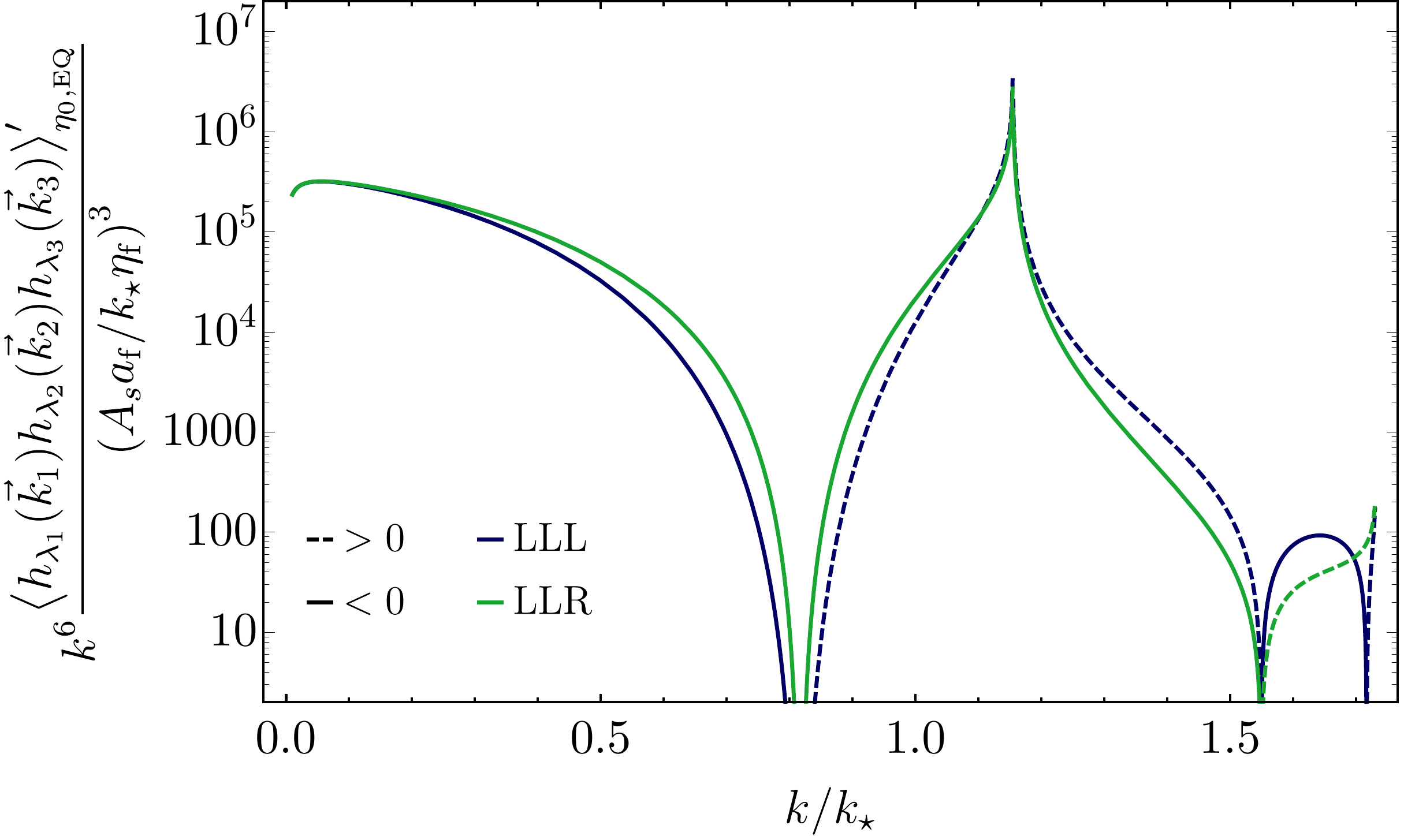}
\caption{The normalised  shapes for the various primordial bispectra in the equilateral configurations.}
\label{fig: Bispectrum GW}
\end{figure}

\vskip 0.3cm
\noindent
\paragraph{The primordial bispectrum of GWs.} Since the GW source is non-linear, the three-point correlator of the GWs is not vanishing. Its computation is straightforward in the approximation of Gaussian initial perturbations~\cite{errgw}
\begin{eqnarray}
&&\hspace{-0.5cm}\big< h_{\lambda_1}(\eta,\vk_1) h_{\lambda_2}(\eta,\vk_2) h_{\lambda_3}(\eta,\vk_3) \big>'  \hspace{-0.1cm}= \hspace{-0.1cm}
 \left(\frac{8\pi}{9} \right)^3
  \!\!\int \!\dd^3 p_1
  \frac{1}{k_1^3 k_2^3 k_3^3 \eta^3} \nonumber\\
  & \cdot&
   e^*_{\lambda_1}(\vk_1,\vp_1) e^*_{\lambda_2}(\vk_2,\vp_2) e^*_{\lambda_3}(\vk_3,\vp_3)
  \frac{\Pz(p_1)}{p_1^3} \frac{\Pz(p_2)}{p_2^3} \frac{\Pz(p_3)}{p_3^3} \nonumber\\
  & \cdot&
  \Big[ \Big(\cos(k_1\eta) \Ic\Big( \frac{p_1}{k_1},\frac{p_2}{k_1}\Big) + \sin(k_1\eta) \Is\Big( \frac{p_1}{k_1},\frac{p_2}{k_1}\Big)\Big) \nonumber \\
 & \cdot&(1\rightarrow 2\, {\rm and}\, 2\rightarrow 3)\cdot  (1\rightarrow 3\, {\rm and}\, 2\rightarrow 1)\Big],
\label{bispectrum}
\end{eqnarray}
where $\vec{p}_2 = \vec{p}_1 - \vec{k}_1$,  $\vec{p}_3 = \vec{p}_1  + \vec{k}_3$, and
where $e^*_{\lambda}(\vk,\vp)=e^{*ij}_{\lambda}(\vk)p_ip_j$ are the polarisation tensors and $\lambda=$ L,R.   The bispectrum of GWs is dominated by the equilateral configuration~\cite{long}, $k_1\simeq k_2\simeq k_3\equiv k$, as expected since it is sourced by
gradients of the curvature perturbations when the latter re-enter the horizon.
For the equilateral configuration and monochromatic power spectrum (Eq.~\eqref{shape}), 
the bispectrum today at time $\eta_0$ is
\begin{align}
%&&  \!\!\!\!\!\!\!\!  \!\!\!\!\!\!\!\!
\big< h_{\lambda_1}(\vk_1) h_{\lambda_2}(\vk_2) h_{\lambda_3}(\vk_3) \big>'_{\eta_0,\text{\tiny EQ}}  = \left(\frac{A_s a_{\rm f}}{k^2 k_\star\eta_{\rm f}}\right)^3
\frac{1024\pi^3}{729} \nonumber\\
%&&
\cdot \left |\frac{1}{\sqrt{2}} {\cal I} \left(\frac{k_\star}{k},\frac{k_\star}{k}\right) \right |^3   \frac{\theta(\sqrt{3} k_\star-k)}{\sqrt{3k_\star^2/k^2-1}}{\cal D}_{\lambda_1\lambda_2\lambda_3}\left(\frac{k_\star}{k}\right),
\label{bispectrumd}
\end{align}
where $\eta_{\rm f}$ is a time well after the modes have entered the horizon but before the thermal degrees of freedom change (we have also taken $a_0=1$).
The function ${\cal D}_{\lambda_1\lambda_2\lambda_3}(x)=365/6912-61x^2/192+9x^4/16-x^6/4$ for the RRR and LLL  polarisations and  ${\cal D}_{\lambda_1\lambda_2\lambda_3}(x)=x^6(-4+1/x^2)^2(-12+5/x^2)^2/768$ for the other combinations, see Fig. \ref{fig: Bispectrum GW}.
Note that in Eq.~\eqref{bispectrumd} we have dropped the phases of the bispectrum. This is a crucial point when asking if the bispectrum can be observed by LISA.

\noindent

\noindent
\vskip 0.3cm
\noindent

\paragraph{Tensor non-Gaussianity is not locally observable.}
The bispectrum calculated above is defined over a constant time hypersurface and hence not directly locally observable.
Is tensor non-Gaussianity actually measurable by LISA? As explained previously, the answer is no because the light cone includes signals generated
in a large number ($\sim 10^{40}$) of independent Hubble patches, and hence the observed signal should sum to be highly Gaussian. How this is consistent with the bispectrum calculation has been a cause of some confusion, which we clarify here.

Non-Gaussianity is present when there are correlations between different Fourier modes,
so non-Gaussianity is synonymous with ``phase correlations"~\cite{Matsubara:2006ep}. For example, correlations between three approximately equal wavenumbers forming a positive bispectrum triangle correspond to the signal being relatively more concentrated at real spaces peaks.
 However, at the time and location of observation, the phases are almost completely uncorrelated, as we now explain.

LISA measures the effect of a collection of  GWs arriving at the detector from all possible directions, not a single one. Consider the change in the light-travel time for a photon emitted from ${\vec x}_1$ at one end of the detector at time $\eta_0$ and arriving at the other end of the arm (at relative position $\vec{L}$)  due to a passing gravitational wave. Integrating the strain along the photon path gives~\cite{snr}
\begin{multline}
\Delta\eta(\eta_0)=\frac{L}{2}\int\frac{\d^3k}{(2\pi)^3}\,e^{i \vec{k} \cdot\vec{x}_1}\sum_\lambda e_\lambda(\vec k,\hat L) \\
\cdot\,
\left[e^{ i k \eta_0}h_\lambda(\vec k){\cal M}(\hat L\cdot\hat k,k) + e^{ -i k \eta_0}h^*_\lambda(-\vec k){\cal M}^*(-\hat L\cdot\hat k,k) \right],
\end{multline}
where
\begin{equation}
{\cal M}(\hat L\cdot\hat k,k)=e^{i k L(1-\hat L\cdot \hat k)/2}{\rm sinc}\left[\frac{k L(1-\hat L\cdot \hat k)}{2}\right]\quad
\end{equation}
(summing the time for the light to come back to the initial point does not change the argument). The bispectrum of the $\Delta\eta$ measured in three arms at locations $\vec x_i$ will be proportional to objects like
\begin{multline}
\!\!\prod_{i=1}^{3}\int \d^3 k_i \,e^{i\vec x_i \cdot \vec k_i}e^{\pm i k_i \eta_0}{\cal M}(\hat L_i\cdot\hat k_i,k_i)\\
\cdot\,\big< h_{\lambda_1}(\vk_1) h_{\lambda_2}(\vk_2) h_{\lambda_3}(\vk_3) \big>.
\end{multline}
Since the bispectrum is peaked around some momentum $k_\star\gg \eta_0^{-1}$, and ${\cal M}$  varies slowly relative to the rapidly oscillating terms
$\exp (i \sum_i \pm k_i\eta_0)$,  these integrals average to zero except where $\sum_i \pm k_i=0$ because of the delta function in the three-point correlation. Note that the power spectrum is not affected since the corresponding phases cancel.

The non-zero contribution from $\sum_i \pm k_i=0$, corresponding to all three $\vec k$ wave vectors being aligned, is the in-principle observable signal from correlating three wavelengths emitted by the same Hubble patch in a particular direction. However, the alignment of directions for $\sum_i \vec k_i=0$ to imply $|\sum_i \pm k_i\eta_0| \alt 1$ has to be very precise, with angular precision $\delta\theta \sim 1/\sqrt{k\eta_0}$, and hence contributes to a negligible fraction $\sim (k\eta_0)^{-1} \ll 1$ of the integral over all angles. The number of independent Hubble patches contributing to the signal is $N\propto (k\eta_0)^2$, so the $\sim (k\eta_0)^{-1}\propto 1/\sqrt{N}$ scaling is just the one expected when measuring the bispectrum of the sum of $N$ independent signals.
This is a consequence  of the central limit theorem: within the measurement time, the detector does not measure a single wave, but a sum of GWs within a given momentum width from all directions, so the signal is strongly Gaussianised.

Could the signal be measured by collecting observations over a long observation time $\Delta \eta$? The phases will average if the range of $k$ values included in the observed band is large compared to $1/\eta_0$. In a total observation time $\Delta\eta$, the bandwidth that is in principle resolvable is $\sim 1/\Delta\eta$. However, since $\Delta\eta\ll\eta_0$ for any observation time small compared to the age of the universe, the signal would still be lost.

How about building a large array of LISA-like detectors? In principle a large array could capture the wavefront from each source horizon volume at multiple locations today, giving observable phase correlations. The array would however have to be cosmologically large to capture more than a tiny fraction of the signal, since the correlated shells of GW emission have a radius of $\eta_0$ today. As the array is made larger, the phases also decorrelate for other reasons, for example due to variations in the Shapiro time delay as the waves propagate across the inhomogeneous universe. This randomises the GW phases, and suppresses the GW bispectrum.

To demonstrate  this effect, consider the propagation of the GW through the inhomogeneous universe. We can work in the geometrical optic limit where the wavelength of the GW is much smaller than the size of the gravitational potentials (though micro-lensing events could also be relevant). The Shapiro time delay is given by the integral along the GW path of the potential, with total delay at position $\vec x$ for a GW observed in direction $-\hat k$
\begin{equation}
\delta\eta(\vec x) = 2\int_{\eta_e}^{\eta_0}{\rm d}\eta'\Psi(\vec x + \hat k(\eta'-\eta_0),\eta'),
\end{equation}
where $\eta_e$ is the emission time (which we can take to be zero). When received, the GW has therefore acquired a phase shift of $k\delta\eta$ compared to the propagation in a homogeneous universe. This is not a problem, as long as the phase shift is the same for all the GW measurements (it would just change the overall phase). However, if they vary, the average correlation of waves at three points would pick up a factor of $\langle \exp(i\sum_i k_i\delta\eta_i)\rangle$. Averaging the exponentials assuming Gaussian $\delta\eta_i$ gives an exponential suppression of the bispectrum with a product of terms of the form $e^{-k_ik_jC_{ij}/2}$, where $C_{ij}$ is a correlation of a time delay difference between two of the measurements. For cosmologically separated observation points, the suppression eventually becomes $\sim e^{-\sum_i k_i^2\sigma^2/2}$~\cite{long}, where $\sigma=\langle \delta\eta^2 \rangle^{1/2} \approx 10^{-4}\eta_0$~\cite{hu}, which wipes out the signal since $k\eta_0 \sim 10^{16}$.

The only remaining way that there could be observable non-Gaussianity in the GW distribution is if there were long-range correlations between Hubble patches at the time of GW emission, for example due to squeezed non-Gaussianity of the primordial perturbation modulating the local fluctuation amplitude. This could in principle lead to the observed GW power varying over observably large angular scales. However, if the dark matter is PBH, the abundance of PBH is very sensitive to the amplitude of perturbations, and would vary spatially if there were long-range variations of the local power spectrum amplitude. The observed large-scale homogeneity of the dark matter density (absence of CDM isocurvature modes) therefore also rules out this option at any significant level.
The prediction of a purely Gaussian isotropic gravitational wave background associated with PBH formation is robust. The quadrupole sources in each independent horizon volume also have uncorrelated orientations (unless there is large anisotropic squeezed primordial non-Gaussianity), so the observed sum of the signals from many volumes is also expected to be unpolarised to high accuracy.

\noindent
\vskip 0.3cm
\paragraph{Conclusions.}
\noindent
If    most (if not all) of  the dark matter is composed by PBHs, this is a  very economical option since no physics beyond the Standard Model is required.
If the PBHs forming the dark matter have a mass of the order of $10^{-12} M_\odot$ this scenario is still observationally viable, and also testable since it inevitably produces a background of gravitational waves that would be detectable by LISA.
Although the gravitational wave source is intrinsically non-Gaussian, the observed signal today should be isotropic and Gaussian.
The task of distinguishing the signal from that from phase transitions or inflationary sources (which also generically predict isotropic Gaussian backgrounds) must then rely on detailed study of the power spectrum shape, a topic that deserves further study.
If new more robust  constraints on $f_\PBH$ appear, they can be satisfied by decreasing  $A_s$ by a small amount (since $f_\PBH$ is exponentially sensitive to $A_s$), which could still leave a potentially detectable GW signal associated with a smaller PBH fraction. 
Finally, the GW signal associated with a narrow mass range of PBHs is peaked in frequency, and hence should be distinguishable from a non-primordial stochastic GW background from astrophysical sources characterised by approximately power-law spectra, such as the signal from black hole mergers.

\vskip 0.3cm
\noindent

\paragraph{Acknowledgments.}
\noindent
  We warmly thank the anonymous referee for asking about the propagation effect.  We thank A. Katz for illuminating discussions on the microlensing and neutron star constraints on the PBH abundance and D. Racco for many discussions. We thank C. R. Contaldi for discussions on the GW propagation and M. Hindmarsh for discussion of Gaussianity. N.B. acknowledges partial financial support by the ASI/INAF Agreement I/072/09/0 for the Planck LFI Activity of Phase E2. He also  acknowledges financial support   by  ASI Grant 2016-24-H.0.
A.R. is  supported by the Swiss National Science Foundation (SNSF), project {\sl The Non-Gaussian Universe and Cosmological Symmetries}, project number: 200020-178787. A.L. is supported by the European Research Council under the European Union's Seventh Framework Programme (FP/2007-2013) / ERC Grant Agreement No. [616170].

%\end{acknowledgments}
%\appendix
%\begin{appendices}

%\section{Appendix A: The non-Gaussian statistics }
%%%%%%%%%%%%%%%%%%%%%%%%%%%%%%%%%%
%\setcounter{equation}{0}
%\renewcommand{\theequation}{A.\arabic{equation}}
%\noindent
%\end{appendices}

\bigskip

%\bigskip

\end{document}